\pdfoutput=1
\documentclass{ieeeaccess}
\usepackage{cite}
\usepackage{amsmath,amssymb,amsfonts}
\usepackage{algorithmic}
\usepackage{graphicx}
\usepackage{textcomp}
\usepackage{soul}
\def\BibTeX{{\rm B\kern-.05em{\sc i\kern-.025em b}\kern-.08em
    T\kern-.1667em\lower.7ex\hbox{E}\kern-.125emX}}
\begin{document}
This work has been submitted to the IEEE for possible publication. Copyright may be transferred without notice, after which this version may no longer be accessible.

\history{Date of publication xxxx 00, 0000, date of current version xxxx 00, 0000.}
\doi{10.1109/ACCESS.2017.DOI}

\title{Perpetual Contract NFT as Collateral for DeFi Composability}
\author{\uppercase{Hyoungsung Kim}\authorrefmark{1},
\uppercase{Hyun-sik Kim\authorrefmark{1}, and Yong-Suk Park}.\authorrefmark{1}}

\address[1]{Korea Electronics Technology Institute (KETI), South Korea}

\tfootnote{This research was supported by Culture, Sports and Tourism R\&D Program through the Korea Creative Content Agency grant funded by the Ministry of Culture, Sports and Tourism in 2022 (Project Name: Open Metaverse Asset Platform for Digital Copyrights Management, Project Number: R2022020034, Contribution Rate: 100\%)}

\markboth
{H. Kim \headeretal: Perpetual Contract NFT as Collateral for DeFi Composability}
{H. Kim \headeretal: Perpetual Contract NFT as Collateral for DeFi Composability}

\corresp{Corresponding author: Yong-Suk Park (e-mail: yspark@keti.re.kr).}

\begin{abstract}

Ethereum and its standardized token interface have formed decentralized finance (DeFi), an open financial system based on blockchain smart contracts. The DeFi ecosystem has become richer with the introduction of DeFi composability projects, such as Lido finance and Curve finance. DeFi composability denotes the concatenation of DeFi services in which each DeFi service locks assets as collateral and gives another asset as liquidity of locked assets to providers. Providers use the tokens given for other concatenated DeFi services, such as lending, decentralized exchanges (DEXs), and derivatives. The DeFi ecosystem uses ERC-20 tokens which can represent the value of an asset. ERC-721 non-fungible tokens (NFTs) are not widely adopted in DeFi, since they represent rights to an asset and are not considered appropriate for valuation. In this paper, we propose a new concept, perpetual contract NFT, which exploits perpetual future contracts in the cryptocurrency derivatives market. Unlike futures contracts in a traditional derivatives market, in the cryptocurrency derivatives market, most futures contracts are perpetual. In addition, the value of futures contract is backed by collateral. Therefore, if we mint the rights to perpetual contracts as NFT, we can use the perpetual contract NFT as collateral for DeFi composability. To validate our proposal and its profitability, we experiment with the position NFT of Uniswap v3. Through validation, we show that our concept works in real-world scenarios.

\end{abstract}


\titlepgskip=-15pt

\maketitle

\section{Introduction}
\label{sec:introduction}
Since Satoshi Nakamoto proposed Bitcoin as a cryptocurrency for peer-to-peer (P2P) electronic cash systems \cite{b1}, there have been many tries to use cryptocurrency as cash alternative in payment systems. Nevertheless, these tries were unsuccessful because of the high fluctuation of cryptocurrency prices. However, the advent of Ethereum \cite{b2} and its standardized token interface, such as ERC-20 and ERC-721, opened the era of decentralized finance (DeFi). Specifically, the standardized token of Ethereum enables diverse DeFi services, such as lending, decentralized exchanges (DEXs), derivatives, etc. DeFi services use a standardized token pegged to a fiat currency, such as the US dollar, to deal with the high fluctuation of cryptocurrency prices; this pegging token is named stablecoin. Strictly speaking, the term "stablecoin" for the pegging "token" is a misnomer. However, in the cryptocurrency industry, this term is widely accepted and used instead of "stable token".

Cryptocurrency lending is one of the notable DeFi services in which cryptocurrency holders provide their assets to a lending platform to earn lending fees. Popular lending service protocols include MakerDAO \cite{b3} and Aave \cite{b4}. Furthermore, a provider can collateralize their provided asset for borrowing up to a max loan to value (LTV). For example, assume the LTV of the cryptocurrency of Ethereum, Ether, is 80\%. Under this LTV, if Ether is provided to the lending platform, the lending platform pays back lending fees in return. In addition, for liquidity of the provided asset, the provider can set the asset as collateral and get a loan equivalent to 80\% price of the collateral from the lending platform. However, if the value of the collateral decreases below certain threshold, the lending platform liquidates the collateral to recover the loan.

In Ethereum, there are two types of standard token interfaces: ERC-20 \cite{b5} and ERC-721 \cite{b6}. ERC-20 tokens represent the asset value, such as the value of an asset relative to a fiat currency. ERC-721 tokens, also known as non-fungible tokens (NFTs), represent the rights associated with the asset. For example, if an artist mints a piece of work as an NFT, the NFT denotes the work’s digital title deed. ERC-20 tokens are suitable to be used as collateral to set LTV rates. Hence, ERC-20 makes lending and liquidating process possible due to its value representing characteristic.

For example, let us assume the use of wrapped Ether (WETH) in a DeFi lending service. WETH is an ERC-20 token pegged to Ether, each WETH representing a value of Ether \cite{b7}, so a WETH holder can exchange WETH for Ether at almost the same market rate. Therefore, in a DeFi lending platform, the WETH holder can borrow assets using WETH as collateral. The DeFi lending platform sets the LTV of collateralized WETH to handle the fluctuation of WETH price. When the WETH price decreases steeply, a DeFi lending platform liquidates collateralized WETH to recover costs. On the other hand, ERC-721, which is used for digital collection NFTs \cite{b8}, \cite{b9}, does not have clear criteria for valuation. If someone paid one WETH for an NFT representing the rights to a digital collection, there is no guarantee that it will retain the one WETH value in the future. Hence, NFTs cannot be used as primary collateral in DeFi lending platforms. However, in this paper, we present a new concept to utilize the characteristic of ERC-721 as collateral by exploiting a perpetual contract of futures in the cryptocurrency derivatives exchanges.

The remainder of this paper is organized as follows. Section II, as background, introduces perpetual contracts, one of the futures contracts in cryptocurrency derivatives exchanges, to help understand the perpetual contract NFT. In addition, we present DeFi composability and its use case. The use case shows how DeFi composability and perpetual contract NFT can be used as trading strategies for the futures market. Section III presents a more specific use case scenario with Uniswap v3 \cite{b10} to validate profitability. Section IV summarizes our work and concludes the paper.

\section{Perpetual contract NFT and use case scenarios}
In this section, to help understand the concept of perpetual contract NFT, we introduce perpetual contracts specialized in futures trading of cryptocurrency derivatives exchanges. After the introduction to perpetual contracts, we present perpetual contract NFTs and demonstrate a use case where it is used as collateral to maximize profit or hedge liquidation in futures trading.

\subsection{Perpetual contracts}
In a traditional derivatives market, traders agree to buy or sell some underlying asset, such as gold or oil, at a predetermined price and time in the future through financial futures contract. The transaction of the asset must take place at the set price at the contract expiration date. However, in the cryptocurrency derivatives market, most futures contracts do not have a predetermined expiration date; these contracts are called perpetual contracts. Perpetual contracts allow traders to deposit cryptocurrencies like Bitcoin, Ether, or stablecoins as collateral to derivatives exchanges; using these collaterals, traders make financial futures contracts on select cryptocurrencies. Furthermore, traders can settle perpetual contracts at any time to claim profits or avoid liquidation.

For example, in a perpetual contract for a derivative of increasing Ether price, known as a long position, perpetual contract holders can claim profit at any time or hold the position as long as they want because there is no predetermined contract expiration date. However, if the derivatives price moves differently from the trader’s prediction, the wrong prediction reduces the value of the perpetual contract. When the trader’s collateral is not enough to back the damaged perpetual contract, the exchange liquidates the trader’s collateral to recover losses. To avoid losing all collateral, traders must settle their perpetual contracts before liquidation.

\subsection{perpetual contract NFT}
Collateral backs the value of the perpetual contract, so the perpetual contract holder can settle the contract to claim their collateral with profit or loss whenever they want. That is, each perpetual contract denotes the right to claim the collateral. Therefore, we can mint the perpetual contract as an NFT representing the rights to the collateral. For example, assume a trader makes a perpetual contract using 100 US dollars stablecoin as collateral for futures, and the market gives a perpetual contract NFT to the trader. At the time of making the NFT, the NFT holder can settle the contract to claim 100 US dollars stablecoin because holding the NFT denotes that the holder has the right to claim the collateral. Hence, a perpetual contract NFT can be seen as possessing the value of the collateral, since it represents the right to claim the collateral. As a result, the perpetual contract NFT holder can utilize it as collateral in other DeFi services, such as lending, DEXs, or derivatives.

In the following subsection, we present use cases of perpetual contract NFT for DeFi composability in the futures market. In addition, in the next section, we propose a scenario to utilize a position NFT of Uniswap v3 \cite{b10}, a popular DEX, as collateral.

\Figure[t!](topskip=0pt, botskip=0pt, midskip=0pt)[width=\textwidth]{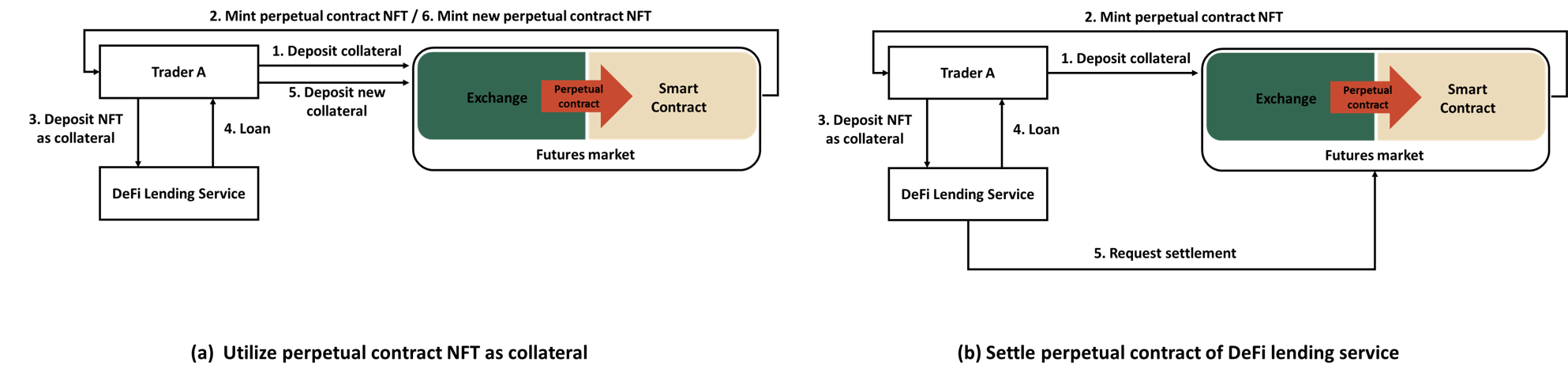}
{\textbf{Use case of perpetual contract NFT.} {\fontfamily{Calibri}\selectfont (a)} denotes {\fontfamily{Calibri}\selectfont Trader A} utilizing  perpetual NFT as collateral to maximize profit or hedge liquidation. {\fontfamily{Calibri}\selectfont (b)} denotes {\fontfamily{Calibri}\selectfont DeFi lending service} requesting liquidation of the collateral by {\fontfamily{Calibri}\selectfont Trader A}.
\label{fig1}}

\subsection{Use case}
\label{use case}
Valuation of ordinary NFTs, such as digital collections, game items, and lands in virtual worlds, depends on a previous selling price; namely, there are no objective criteria for valuation. Hence, these NFTs are not recommended to be used as collateral in DeFi services. However, the value of perpetual contract NFT is backed by collateral, so it is possible to exploit the valuation of perpetual contract NFT as collateral in various DeFi services. In addition, perpetual contract NFT follows standard interface ERC-721, so it is possible to adopt it in trading scenarios, such as peer-to-peer (P2P), peer-to-exchange (P2E), and exchange-to-exchange (E2E). Hence, perpetual contract NFT can be adopted to DeFi composability. DeFi composability, also known as DeFi Lego, is the concatenation of DeFi services to maximize profit by giving liquidity to a locked asset. This subsection presents additional background details about DeFi composability and demonstrates a practical use case of DeFi composability and perpetual contract NFT as strategies to maximize profit and avoid liquidation in the futures market.

\subsubsection{DeFi composability}
DeFi composability, also called DeFi Lego, is a way to maximize profit by giving liquidity to a locked asset. Liquidity describes the degree to which an asset can be quickly converted into ready cash, maintaining its market price value. For example, Beacon chain \cite{b14} is testing a new consensus algorithm for Ethereum 2.0. To be a miner, who can get block generation rewards \cite{b15} from Beacon chain, they have to stake 32 Ether as a deposit, and Beacon chain locks this deposit until Ethereum 2.0 is released. Even though miners get block generation rewards as incentive for staking, they lose the liquidity for 32 Ethers. Hence, Lido finance \cite{b13} released a DeFi staking delegation service. Lido finance gives delegators a block generation reward and liquidity for locked Ether. Specifically, to give liquidity to delegated Ether, when Ether holders hand over their Ethers to Lido finance, Lido finance gives the same amount of stETH, a tokenized version of staked ether, as delegated Ether. Namely, Ether backs worth of stETH like collateral. As a result, stETH holders can use it like Ether in other DeFi services, such as Curve finance \cite{b16}, and stETH holders get a block generation fee for staking to Beacon chain from Lido finance. Therefore, unlike direct staking to the Beacon chain, the staking delegation of Lido finance provides both liquidity of staked Ethers and block generation rewards in the Beacon chain.

Delegated Ethers to Lido finance are locked in the Beacon chain, so it needs a service to swap stETH to Ether. For this service, Curve finance \cite{b16} provides the stETH swap. In Curve finance, stETH or Ether holders can provide their assets to a liquidity pool (LP), which helps asset swap; as rewards for providing liquidity, providers get an LP token and swap fee. Namely, concatenated DeFi services, such as Lido finance and Curve finance, give Ether holders cumulative rewards: Beacon chain rewards, swap fees, and LP tokens. In addition, they can utilize LP tokens for other DeFi services to get more profit. These concatenated DeFi services for cumulative rewards are called DeFi composability.

\subsubsection{Perpetual contract NFT as collateral for futures}
Perpetual contract NFT is backed by collateral, so valuation is possible. As a result, perpetual contract NFT can be utilized as collateral for DeFi composability like ERC-20 tokens. Uniswap v3 \cite{b10} position NFT shows a possibility of perpetual contract NFT. In Uniswap v3, liquidity providers concentrate their assets to a spot of LP as liquidity and get a position NFT representing the right of liquidity in the spot. There is no predetermined expiration date of liquidity, so liquidity providers can reduce liquidity in LP whenever they want. Therefore, the position NFT is a kind of perpetual contract NFT. If liquidity providers exploit this position NFT as collateral for DeFi composability, such as a loan, they can make more profit without rebalancing liquidity. We discuss in more detail Uniswap v3 and its position NFT in the next section.

Traders can use perpetual contracts NFT as collateral to maximize profit in the cryptocurrency futures market. Fig. \ref{fig1} shows two cases of perpetual contract NFT to exploit as collateral for loans. Fig. \ref{fig1} {\fontfamily{Calibri}\selectfont (a)} presents the borrowing process to maximize profit. {\fontfamily{Calibri}\selectfont Trader A} makes a perpetual contract of futures like Ether price position using their asset as collateral. After making the perpetual contract, the futures market gives a perpetual contract NFT, representing the rights of the perpetual contract and collateral. To maximize profit, {\fontfamily{Calibri}\selectfont Trader A} delegates these rights to the DeFi lending service by giving NFT for a loan; using this loan, {\fontfamily{Calibri}\selectfont Trader A} makes a new perpetual contract. When {\fontfamily{Calibri}\selectfont Trader A} wants to close all perpetual contracts to claim profits and collateral, {\fontfamily{Calibri}\selectfont Trader A} settles the second perpetual contract NFT and repays for the loan to the DeFi lending service to take back the perpetual contract NFT. Next, {\fontfamily{Calibri}\selectfont Trader A} settles this perpetual contract NFT to take back their collateral and profits. Delegating the right of collateral to DeFi lending service denotes that DeFi lending service can claim collateral of {\fontfamily{Calibri}\selectfont Trader A} to futures market for recovering loan. Fig. \ref{fig1} {\fontfamily{Calibri}\selectfont (b)} shows recovering by the DeFi lending service. When the value of a perpetual contract is down, the DeFi lending service request settlement before the futures market liquidate traders’ collateral and gets traders’ collateral to recover the loan.

In addition, by using perpetual contract NFT as collateral, traders can avoid liquidation without rebalancing their position. In the derivatives market, futures prices do not move linearly, so even if a trader makes an accurate long-term price prediction, liquidation may occur due to short-term price movements. Therefore, traders reserve assets to respond, such as price movements, or rebalance their contracts to avoid liquidation. However, perpetual contract NFT helps to let traders exploit all their assets for trading futures. For example, assume a trader makes a perpetual contract as a long position for futures price using all assets, but the short-term price is decreasing. In this situation, the market will liquidate the trader’s collateral; if there is a perpetual contract NFT, the trader exploits it as collateral to borrow assets from the lending service to make a new position. With this new position, the trader can make profit for an additional margin to avoid liquidation without rebalancing the position.

\section{Profitability validation}
In this section, an overview of automated market maker (AMM) and concentrated liquidity is given as background. AMM is the underlying protocol of Uniswap \cite{b17}. Concentrated liquidity is a mechanism which was aggregated to AMM in the third version release of Uniswap, Uniswap v3 \cite{b10}. In addition, we validate the profitability of NFT as collateral using concentrated liquidity of Uniswap v3.

\subsection{background}
Uniswap proposed a simple AMM protocol for DEX, and Uniswap v3 applied concentrated liquidity to AMM. These contributions brought forth improvements of other DEXs, such as Curve finance \cite{b16}. In this subsection, we introduce the concepts behind AMM and concentrated liquidity in Uniswap, and we use concentrated liquidity to show profitability of NFT as collateral.

\subsubsection{Automated market maker}
Unlike centralized exchange (CEX) that uses an order book system for trading, DEX, such as Uniswap, uses AMM for trading. In the order book trading of CEXs, centralized intermediaries match prices through bid and ask to help trading. However, in the AMM trading of DEX, a decentralized automated asset swapping algorithm, known as AMM, makes trading possible. The AMM adjusts a swap ratio of asset pairs in a liquidity pool (LP) for automated algorithmic swaps, and liquidity providers present assets as liquidity to the LP to get a swap fee. Furthermore, LPs in Uniswap consist of the liquidity provider’s asset pairs. For example, UNI/WETH LP consists of two asset pairs: a governance token of Uniswap (UNI) and wrapped Ether (WETH); the AMM helps traders swap UNI with WETH and vice versa by adjusting the swap ratio of this pair. The AMM of Uniswap, known as the constant product formula (CPF) is:

$$x \cdot y = k$$

\noindent
where $x$ and $y$ denote the respective reserves of two assets X and Y, and $k$ is constant  \cite{b17}. Fig. \ref{fig2} {\fontfamily{Calibri}\selectfont (a)} shows the AMM curve of the WETH-UNI pair.

\Figure[t!](topskip=0pt, botskip=0pt, midskip=0pt)[width=0.85\linewidth]{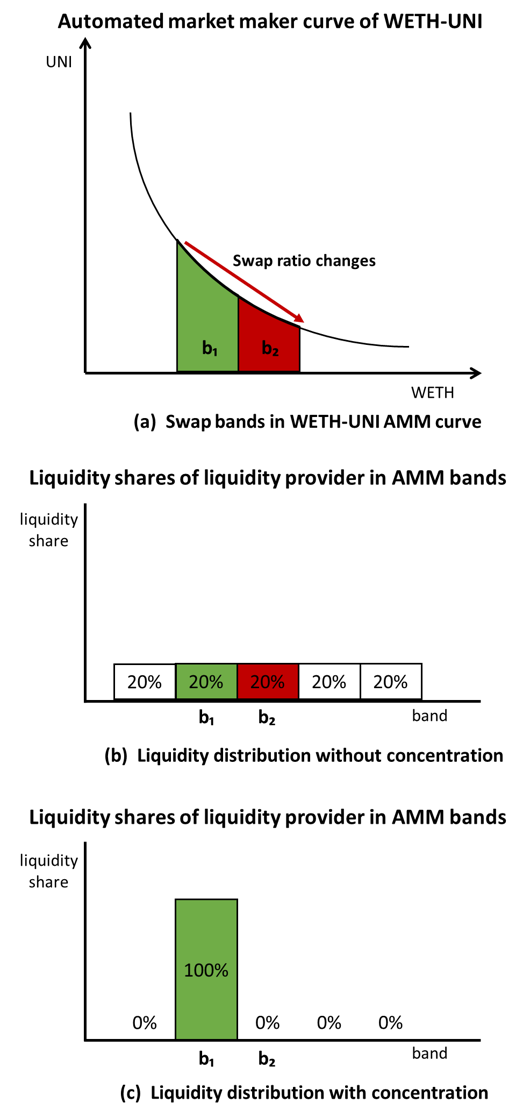}
{\textbf{Uniswap AMM curve of WETH-UNI pair and liquidity distributions}  \label{fig2}}

\subsubsection{Concentrated liquidity of Uniswap v3}
Prior to Uniswap v3 release, previous versions of Uniswap allocated liquidity evenly, forming a uniform distribution. In Uniswap, swap fee rewards depend on liquidity shares. If liquidity is distributed evenly, the liquidity provider may miss out on swap fee rewards.

For example, assume all swaps happen in the band {\fontfamily{Calibri}\selectfont b\textsubscript{1}} and that there is a liquidity provider who has enough assets to dominate liquidity shares of a specific band of AMM. If the liquidity provider concentrates all assets to a band, as in Fig. \ref{fig2} {\fontfamily{Calibri}\selectfont (c)}, the liquidity provider achieves 100\% of the liquidity shares of the band and takes 100\% of the collected swap fees in {\fontfamily{Calibri}\selectfont b\textsubscript{1}}. However, in evenly distributed liquidity, liquidity is allocated to all bands regardless of the swap’s existence. As a result, even if the liquidity provider has enough assets to dominate band {\fontfamily{Calibri}\selectfont b\textsubscript{1}}, the liquidity provider cannot use all assets to dominate the liquidity share of the band. Instead, only swap fees for the 20\% of liquidity allocated in band {\fontfamily{Calibri}\selectfont b\textsubscript{1}} can be collected.

Unlike previous versions of Uniswap, in Uniswap v3, liquidity providers allocate their liquidity manually. Liquidity providers can concentrate in a specific band in AMM to maximize their swap fee rewards as shown in Fig. \ref{fig2} {\fontfamily{Calibri}\selectfont (c)}. The liquidity provider dominates the liquidity shares of the band and takes all collected swap fees of the band. Concentrated liquidity can exploit all assets to a band to maximize profit, but it does not generate profit when the swap ratio changes. The swap ratio changes when swaps happen more frequently for one asset than another. Fig. \ref{fig2} {\fontfamily{Calibri}\selectfont (a)} illustrates swap ratio changes in AMM. If liquidity allocation is uniformly distributed as in Fig. \ref{fig2} {\fontfamily{Calibri}\selectfont (b)}, liquidity providers can receive swap fee rewards regardless of swap ratio changes. However, if liquidity allocation is concentrated as in Fig. \ref{fig2} {\fontfamily{Calibri}\selectfont (c)} and swap ratio changes from band {\fontfamily{Calibri}\selectfont b\textsubscript{1}} to {\fontfamily{Calibri}\selectfont b\textsubscript{2}}, liquidity providers cannot receive any swap fee since there are no liquidity shares allocated in {\fontfamily{Calibri}\selectfont b\textsubscript{2}}.

For concentrated liquidity, Uniswap v3 presents LP with three types of fee tiers $f  := \{0.05\%, 0.30\%, 1\%\}$ ; additional fee tiers can be enabled by UNI governance \cite{b10}. We can estimate the expected collected fee rewards of the liquidity provider $j$, who concentrates liquidity in the band $i$ with fee tier $x$ as follows:

\begin{equation}
\label{eqn:swap fee reward of bnad}
R_i^j := \frac{L_i^j}{\sum{L_i}} \cdot A_i^x \cdot f_x
\end{equation}

\noindent
where $L_i^j$ denotes liquidity of the liquidity provider $j$ in band $i$, $A_i^x$ denotes accumulated trade volume of band $i$ for 24 hours in LP with fee tier $f_x$, and $f_x \in f$. $L_i^j/\sum{L_i}$ denotes the liquidity shares of $j$ in band $i$, and $A_i^x \cdot f_x$ means unclaimed fees in LP of fee tier $f_x$ for 24 hours. In addition, for $n, m \in \mathbb{N}$, LP of $n$ bands is $LP := \{L_1, L_2, ..., L_n\}$ and liquidity band $i$ consisting of $m$ providers is $L_i := \{L_i^1, L_i^2, ..., L_i^m\}$. We use the above definitions for profitability validation.


\subsection{Profitability of NFT as collateral}
The concentrated liquidity of Uniswap v3 allows liquidity providers to manually allocate liquidity position to maximize profit. However, manual allocation raises concerns about rebalancing the position. When the swap ratio is changed, liquidity providers need to rebalance liquidity distribution to avoid losses. As a result, concerned liquidity providers prefer Uniswap v2 over Uniswap v3. They believe that without optimal position rebalancing, Uniswap v3 gives fewer returns than Uniswap v2 \cite{b18}. In this subsection, we propose a method which will motivate liquidity providers to migrate to Uniswap v3. By using Uniswap position NFT as collateral, liquidity position becomes the futures contract of the derivatives market.


\subsubsection{Uniswap position NFT as collateral}
Fig. \ref{fig3} shows the position NFT of the Uniswap v3 of WETH-UNI with a 0.3\% fee tier LP. When liquidity providers allocate assets to a specific band of LP as liquidity position, liquidity providers get a position NFT representing their rights to the liquidity position in an AMM. We can consider the position NFT as a perpetual contract NFT because the position of Uniswap has similar characteristics to the perpetual contract of the derivatives market. Namely, liquidity providers can hold their position indefinitely and close their position to claim assets whenever they want. Furthermore, reserved assets of liquidity providers back the value of the position, like collaterals backing the value of perpetual contracts. Therefore, position NFT holders get rights to the reserved assets and can claim the assets in that position. In addition, NFT is a standard interface, so it is possible to trade the position NFT in an NFT market. The buyer of a position NFT gets the right to claim the assets bound to the position.

Position NFT holders can make use of DeFi composability to exploit position NFT as collateral to reduce concerns associated with position rebalancing. The value of the position NFT can be assessed by the assets in its position. Based on this valuation, the position NFT holder can borrow additional assets from other DeFi lending services with the position NFT as collateral. Using these assets, holders can make a new position in Uniswap v3 instead of rebalancing their position, or they can use the assets for other DeFi services.

\Figure[t!](topskip=0pt, botskip=0pt, midskip=0pt)[width=0.5\linewidth]{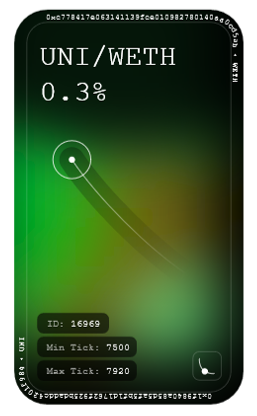}
{\textbf{Liquidity position NFT of WETH-UNI pool in Uniswap v3}  \label{fig3}}

\subsubsection{Profitability validation}
In Uniswap AMM, when trade volumes move from bands $i$ to $i + 1$, liquidity providers must decide whether to rebalance the position or not. Rebalancing raises concerns about losing profit. To mitigate the concerns, we propose the use of position NFT as collateral for DeFi composability, such as lending service, to borrow additional assets. To estimate the profitability from borrowing, assume that there is no additional liquidity except borrowing from provider $j$, and fee tier $f_x$ is fixed until the provider reduces all liquidity. Under these assumptions, we can estimate the borrowing profitability as follows:

\begin{equation}
\label{eqn:profitability}
\frac{L_i^j}{\sum{L_i}} \cdot A_i^x \cdot f_x \le \frac{L_i^j}{\sum{L_i}} \cdot A_i^x \cdot f_x + \frac{L_{i+1}^j}{\sum{L_{i+1}}} \cdot A_{i+1}^x \cdot f_x - L_{i+1}^j \cdot I
\end{equation}

and

\begin{equation}
\label{eqn:borrow asset} 
L_{i+1}^j = LTV \cdot L_i^j
\end{equation}

\noindent
$I$ denotes daily loan interest rate of lending services, and $LTV$ denotes loan to value. For example, when the price of a position NFT is \$100 and LTV rate is 80\%, the position NFT holder can borrow up to \$80 with position NFT as collateral. By rearranging inequality (\ref{eqn:profitability}), we obtain:

\begin{equation}
\label{eqn:rearranged profitability}
 I \le \frac{A_{i+1}^x}{\sum{L_{i+1}}} \cdot f_x 
\end{equation}

Equation (\ref{eqn:rearranged profitability}) shows that profits depend on $A_{i+1}^x$. Let us assume that a liquidity provider $j$ borrows the same amount of UNI and WETH from AAVE [4] with a variable borrow rate for short-term from an LP with fee tier $f_x = 0.30\%$. Fig. \ref{fig4} presents the borrow rate of two assets for 2022 Q1 in AAVE. The average of each asset’s annual variable rate for 1Q is 0.519\% and 0.733\%. Therefore, the annual average borrow rate for the total amount of assets is 0.626\% when a liquidity provider borrows the same volume of each asset. Since Uniswap gives daily fees to liquidity providers \cite{b18}, we can estimate profitability as follows:

\begin{equation}
\label{eqn:estimated profit}
 \frac{0.626}{365} \le \frac{A_{i+1}^x}{\sum{L_{i+1}}} \cdot 0.30
\end{equation}

Therefore, from the futures contract perspective, if provider $j$ expects trade volume in band $i + 1$ to be more than 0.5\% of liquidity in band $i + 1$, provider $j$ will use the position NFT as collateral for a loan to generate more profit. Since using the position NFT as collateral for a loan gives more returns than non\-rebalancing, it offers new alternatives for liquidity providers who are still staying in Uniswap v2 to migrate to Uniswap v3.

\Figure[t!](topskip=0pt, botskip=0pt, midskip=0pt)[width=0.85\linewidth]{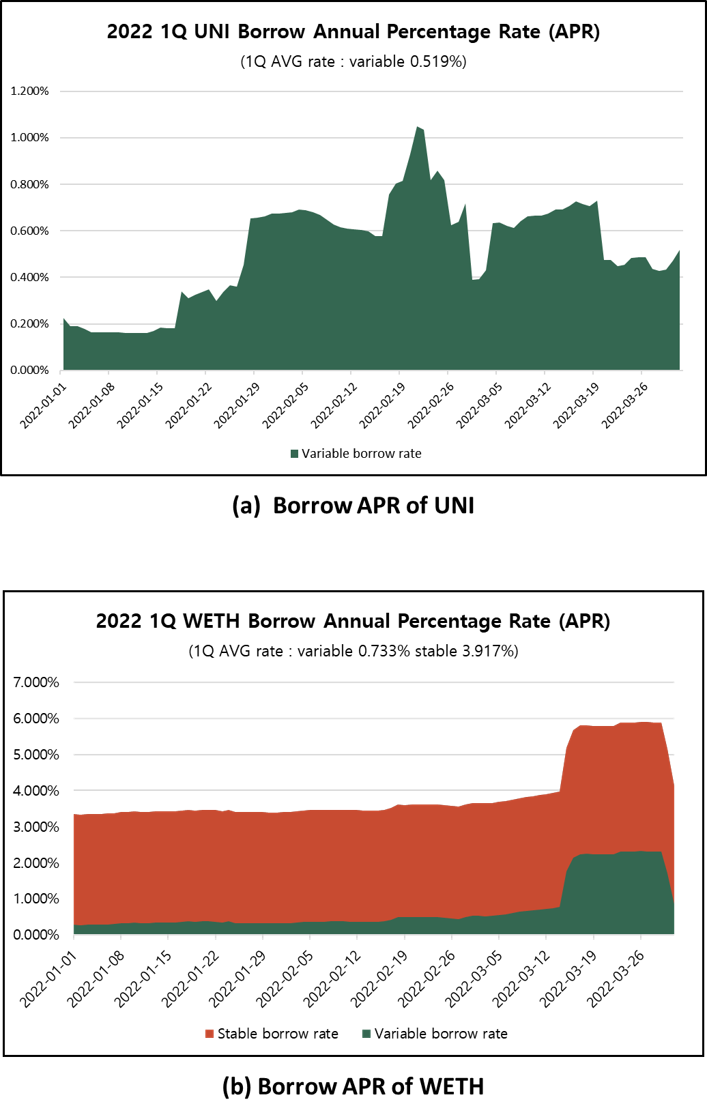}
{\textbf{Borrow Annual Percentage Rate (APR) of UNI and WETH in AAVE for 2022 1Q.} AAVE provides loan services based on collateral. For UNI, only variable borrow rate is available for loan. For WETH, variable and stable rates are available for loan. Data was extracted from a subgraph in AAVE \underline{https://github.com/aave/protocol-subgraphs}. \label{fig4}}

\subsection{Discussion}
In this section, we used the position NFT to provide an example of perpetual contract NFT and prove its profitability. Unlike the traditional financial market, most of the future contracts in the cryptocurrency market do not have a fixed deadline. Such contracts with no expiration dates are called perpetual contracts. We introduced a new perspective for utilizing perpetual contracts through perpetual contract NFT. Furthermore, we validated profitability in the previous subsection which will motivate liquidity provision in Uniswap v3. In our profitability validation process, only short-term loans were considered. For long-term loans, additional considerations are necessary to maximize profit. First, there needs to be a clear understanding that there are many factors that can influence borrow annual percentage rate (APR), such as macroeconomics and serial liquidation in DeFi. Second, impermanent loss may occur during long-term borrowing due to cryptocurrency price fluctuation.


\section{Conclusion}
In this work, we introduced a new perspective for utilizing perpetual contracts as NFT for DeFi composability, such as lending, DEX, and derivatives. Specifically, we proposed minting a perpetual contract as NFT to exploit it in lending services as collateral. Using this lending service, perpetual contract NFT holders can maximize their profit or avoid liquidations in the futures market. A real-world use case was provided through concentrated liquidity and position NFTs used in Uniswap v3. The position NFT in DeFi composability was used to maximize the liquidity provider’s profit without rebalancing by considering the liquidity position in the futures contract.

\EOD

\end{document}